%% file: ow_sysid2015.tex
\documentclass{ifacconf}

\usepackage{graphicx}      
\usepackage[authoryear]{natbib}        
\usepackage{amssymb}
\usepackage{xargs}
\usepackage{ifthen}
\usepackage{bbm}
\usepackage{aliascnt}
\usepackage{stmaryrd}
\usepackage{mathtools}
\usepackage{todonotes}
\usepackage{enumerate}
\usepackage{ushort}
\usepackage{algpseudocode}
\usepackage{algorithm}
\usepackage{subcaption}
\usepackage{ushort}

\input{ow_sysid2015_def}
\begin{document}
\begin{frontmatter}

\title{An efficient particle-based online EM algorithm for general state-space models\thanksref{footnoteinfo}} 

\thanks[footnoteinfo]{This work is supported by the Swedish Research Council, Grant 2011-5577.}

\author[First]{Jimmy Olsson} 
\author[Second]{Johan Westerborn} 

\address[First]{KTH Royal Institute of Technology (e-mail: jimmyol@kth.se).}
\address[Second]{KTH Royal Institute of Technology (e-mail: johawes@kth.se).}

\begin{abstract}                
Estimating the parameters of general state-space models is a topic of importance for many scientific and engineering disciplines. In this paper we present an online parameter estimation algorithm obtained by casting our recently proposed particle-based, rapid incremental smoother (PaRIS) into the framework of online expectation-maximization (EM) for state-space models proposed by~\cite{cappe:2009}. Previous such particle-based implementations of online EM suffer typically from either the well-known degeneracy of the genealogical particle paths or a quadratic complexity in the number of particles. However, by using the computationally efficient and numerically stable PaRIS algorithm for estimating smoothed expectations of time-averaged sufficient statistics of the model we obtain a fast algorithm with very limited memory requirements and a computational complexity that grows only linearly with the number of particles. The efficiency of the algorithm is illustrated in a simulation study. 
\end{abstract}


\end{frontmatter}
\section{Introduction}
This paper deals with the problem of \emph{online parameter estimation} in general \emph{state-space models} (SSM) using \emph{sequential Monte Carlo} (SMC) methods and an \emph{expectation-maximization} (EM) algorithm. SSMs, which are also referred to as \emph{general hidden Markov models}, are currently applied within a wide range of engineering and scientific disciplines; see e.g. \citet[Chapter~1]{cappe:moulines:ryden:2005} and the references therein. In its most basic form \citep[as proposed by][]{dempster:laird:rubin:1977}, the EM algorithm, which is widely used for estimating model parameters in SSMs, is an offline algorithm in the sense that every recursive parameter update requires the processing of a given batch of data. When the batch is very large or when data is received only gradually in a stream, this approach may be slow and even impractical. In such a case, using an \emph{online EM-algorithm} is attractive since it generates a sequence of parameter converging towards the true parameter by processing recursively the data in a \emph{single sweep}.

The algorithm we propose is a hybrid of the online EM algorithm proposed by~\cite{cappe:2009} and the efficient particle-based online smoothing algorithm suggested recently by~\cite{olsson:westerborn:2014:2}. On the contrary to previous algorithms of the same type \citep[see e.g.][]{delmoral:doucet:singh:2009}, which have a quadratic computational complexity in the number of particles, our algorithm stays numerically stable for a complexity that grows only linearly with the number of particles. 

\section{Preliminaries}

We will always assume that all distributions admit densities with respect to suitable dominating measures and we will also assume that all functions are bounded and measurable. 

In SSMs, an unobservable Markov chain $\{ X_t \}_{t \in \nset}$ (the \emph{state process}), taking values in some space $\set{X}$ and having transition density and initial distribution $\hd{\theta}$ and $\Xinit$, respectively, is only partially observed through an observation process $\{ Y_t \}_{t \in \nset}$ taking values in $\set{Y}$. Conditionally on state process, the observations are assumed to be independent with conditional distribution $\md{\theta}$ of each $Y_t$ depending on the corresponding state $X_t$ only; i.e., 
$$
\prob_\theta(Y_t \in \set{A} \mid X_{0:t}) = \int_{\set{A}} \md{\theta}(X_t, y) \, \rmd y \eqsp, 
$$
for all (measurable) $\set{A} \subseteq \set{X}$. The densities above are indexed by a parameter vector $\theta$ that determines completely the dynamics of the model. 

Given a sequence of observations $y_{0:t}$, the likelihood of $\theta$ is given by
\begin{multline*}
L_{\theta}(y_{0:T}) = \\Ê\int \md{\theta}(x_0, y_0) \, \Xinit(x_0) \prod_{t = 1}^T \md{\theta}(x_t, y_t) \hd{\theta}(x_{t - 1}, x_t) \, \rmd x_{0:T} \eqsp. 
\end{multline*}
Unless we are operating on a linear Gaussian model or a model with a finite state space $\set{X}$, this likelihood is intractable and needs to be approximated. If we wish to infer a subset of the hidden states given the observations, the optimal choice of distribution is the conditional distribution $\post{s:s' \mid T; \theta}$ of $X_{s:s'}$, $(s \leq s')$ given the observations $Y_{0:T}$. This is given by
\begin{multline*}
\post{s:s' \mid T; \theta}(x_{s:s'}) = L_{\theta}^{-1}(y_{0:T}) \iint \md{\theta}(x_0, y_0) \Xinit(x_0) \\ 
\times \prod_{t = 1}^T \md{\theta}(x_t, y_t) \hd{\theta}(x_{t - 1}, x_t) \, \rmd x_{0:s - 1} \, \rmd x_{s' + 1:T} \eqsp.
\end{multline*}

We refer to $\post{T;\theta} = \post{T\mid T;\theta}$ as the \emph{filter distribution} and to $\post{0:T \mid T; \theta}$ as the \emph{joint smoothing distribution}.

The EM algorithm computes the maximum-likelihood estimator using some initial guess $\theta_0$ of the same. It proceeds recursively in two steps. First, in the \emph{E-step}, given some parameter estimate $\theta_i$, it computes the $xˆy$
\emph{intermediate quantity}
\begin{multline*}
    Q(\theta, \theta_i) = 
    \E_{\theta_i} \left[\sum_{t = 0}^T \log \md{\theta}(X_t, Y_t) \mid Y_{0:T} \right] \\
    + \E_{\theta_i} \left[\sum_{t = 0}^{T - 1} \log \hd{\theta}(X_t, X_{t + 1}) \mid Y_{0:T} \right] \eqsp,
\end{multline*}
where $\E_{\theta_{i}}$ denotes the expectation under the dynamics determined by the parameter $\theta_{i}$, and in the second step, the \emph{M-step}, it updates the parameter fit according to $\theta_{i + 1} = \arg \max Q(\theta, \theta_i)$. Under weak assumptions, repeating this procedure produces a sequence of parameter estimates that converges to a stationary point of the likelihood. If the joint distribution of the SSM belongs to an \emph{exponential family}, then the intermediate quantity may be written as
$$
    Q(\theta, \theta_i) = \langle \phi(\theta), \post{0:T \mid T; \theta_i}(\af{T}) \rangle - c(\theta) \eqsp,
$$
where $\langle \cdot, \cdot \rangle$Ê denotes scalar product, $\phi(\theta)$ and $c(\theta)$ are known functions, and $\af{T}$ is a (vector-valued) sufficient statistic of additive form. Given the smoothed sufficient statistic $\post{0:T \mid T; \theta_i}(\af{T})$, the M-step of the algorithm can be typically expressed in a closed form via some update function $\theta_{i + 1} = \Lambda(\post{0:T \mid T; \theta_i}(\af{T}) / T)$. Hence, being able to compute such smoothed expectations is in general crucial when casting SSMs into the framework of EM. 

As mentioned, the sufficient statistic $\af{T}(x_{0:T})$ is of \emph{additive form}, i.e.
$$
    \af{T}(x_{0:T}) = \sum_{t = 0}^{T - 1} \addf{t}(x_{t:t+1}) \eqsp,
$$
where all functions are possibly vector-valued. We denote vector components using superscripts, i.e., $\addf{t}(x_{t:t+1}) = (\sadd{t}{1}(x_{t:t+1}), \ldots, \sadd{t}{\ell}(x_{t:t+1}))$.

\begin{exmp} \label{ex:lg}
Consider the \emph{linear Gaussian} SSM
\[
    \begin{split}
        X_{t + 1} &= a X_t + \sigma_V V_t \eqsp, \\
        Y_t &= X_t + \sigma_U U_t \eqsp,
    \end{split}
    \quad t \in \nset \eqsp,
\]
where $\{ V_t \}_{t \in \nset}$ and $\{ U_t \}_{t \in \nset}$ are mutually independent sequences of independent standard Gaussian variables. The parameters of this model are $\theta = (a, \sigma_V^2, \sigma_U^2)$ and the model belongs to an exponential family with sufficient statistics given by
\begin{align*}
    \sadd{t}{1}(x_{t:t+1}) &= x_t^2 \eqsp, & \sadd{t}{2}(x_{t:t+1}) &= x_t x_{t+1} \eqsp, \\ 
    \sadd{t}{3}(x_{t:t+1}) &= x_{t+1}^2 \eqsp, & \sadd{t}{4}(x_{t:t+1}) &= (y_{t+1} - x_{t+1})^2 \eqsp.
\end{align*}
The M-step update function is then given by
$$
    \Lambda(z_1, z_2, z_3, z_4) = \left( \frac{z_2}{z_3}, z_3 - \frac{z_2^2}{z_1}, z_4 \right) \eqsp.
$$
\end{exmp}

When computing smoothed expectations of additive form it is advantageous to use the \emph{backward decomposition} of the joint smoothing distribution. This decomposition comes from the fact that the state process is, conditionally on the observations, still Markov in the forward as well as backward directions. The \emph{backward kernel} $\bk{\post{t; \theta}; \theta}$, i.e., the distribution of $X_t$ conditionally on $X_{t + 1}$ and $Y_{0:t}$, is given by  
$$
\bk{\post{t;\theta};\theta}(x_{t + 1}, x_t) = \frac{\post{t;\theta}(x_t) \hd{\theta}(x_t, x_{t + 1})}{\int \post{t; \theta}(\tilde{x}_t) \hd{\theta}(\tilde{x}_t, x_{t + 1}) \, \rmd \tilde{x}_t} \eqsp.
$$
Using the backward kernel, the joint smoothing distribution may be written as
$$
\post{0:T \mid T; \theta}(x_{0:T}) = \prod_{t = 0}^{T - 1} \bk{\post{t; \theta}; \theta}(x_{t + 1}, x_t) \post{T; \theta}(x_T) \eqsp.
$$
The backward distribution can also be used effectively when estimating expectations of additive form; indeed, if $\statistic{t}(x_t) = t^{-1} \int s_t(x_{0:t}) \prod_{\ell = 0}^{t - 1} \bk{\post{\ell; \theta}; \theta}(x_{\ell + 1}, x_\ell) \, \rmd x_{0:t - 1}$, then $\post{0:t; \theta}(t^{-1} s_t) = \post{t; \theta}(\statistic{t})$ and 
\begin{multline} \label{eq:statistic:recursion}
    \statistic{t}(x_t) = \int \{ \gamma_t \addf{t - 1}(x_{t - 1:t}) + (1 - \gamma_t) \statistic{t - 1}(x_{t - 1}) \} \\Ê
    \times \bk{\post{t - 1; \theta}; \theta}(x_t, x_{t - 1}) \, \rmd x_{t - 1} \eqsp,
\end{multline}
where $\gamma_t = t^{-1}$. The recursion is initialized by $\statistic{0} \equiv 0$. The fundamental idea of the online EM algorithm is to update sequentially the smoothed sufficient statistics using \eqref{eq:statistic:recursion} and plugging these quantities into the updating procedure $\Lambda$. In practice one uses, rather than $\gamma_t = t^{-1}$, some step size $\gamma_t$ satisfying the regular stochastic approximation requirements $\sum_{t = 0}^\infty \gamma_t = \infty$ and $\sum_{t = 0}^\infty \gamma_t^2 < \infty$. Nevertheless, since the backward kernel involves the filter distributions, the recursion \eqref{eq:statistic:recursion} cannot be computed in a closed form, and we are hence forced to approximate the same. We will use \emph{particle methods} for this purpose. 

\section{Particle Methods}

A \emph{particle filter} updates sequentially, using importance sampling and resampling techniques, a set $\{(\epart{t}{i},\wgt{t}{i})\}_{i = 1}^{\N}$ of particles and associated weights targeting a sequence of distributions. In our case we will use the particle filter to target the filter distribution flow $\{Ê\post{t; \theta} \}_{t \in \nset}$ in the sense that
$$
\post[part]{t; \theta}(\testf) = \sum_{i = 1}^{\N} \frac{\wgt{t}{i}}{\wgtsum{t}} f(\epart{t}{i})  \backsimeq \post{t; \theta}(\testf) \quad \mbox{ as } \N \to \infty,
$$
where $\wgtsum{t} = \sum_{i=1}^{\N} \wgt{t}{i}$. Notice that the parameters and weights depend on the parameters even though this is implicit in the notation. In the \emph{bootstrap particle filter}, the sample $\{(\epart{t}{i},\wgt{t}{i})\}_{i = 1}^{\N}$ is updated by, first, resampling the particles multinomially according to weights proportional to the particle weights, second, propagate the resampled particles  forward in time using the dynamics of the state process space, and, third, assigning the particles weights proportional to the local likelihood of the new observation given the particles. The update, which is detailed in Algorithm~\ref{alg:BS}, will in the following be expressed as $\{ (\epart{t + 1}{i}, \wgt{t + 1}{i}) \}_{i =1}^{\N} \gets \PF(\{(\epart{t + 1}{i}, \wgt{t + 1}{i}) \}_{i = 1}^{\N}; \theta)$.

\begin{algorithm}[htb]
    \caption{Bootstrap particle filter}
    \label{alg:BS}
    \begin{algorithmic}[1]
        \Require Parameters $\theta$ and a weighted particle sample $\{(\epart{t}{i},\wgt{t}{i})\}_{i=1}^{\N}$.
        \For{$i = 1 \to \N$}
            \State $I_{t + 1}^i \sim \probdist(\{\wgt{t}{\ell}\}_{\ell = 1}^{\N})$;
            \State Draw $\epart{t + 1}{i} \sim \hd{\theta}(\epart{t}{I_{t + 1}^i}, x_t) \, \rmd x_t$;
            \State Set $\wgt{t + 1}{i} \gets \md{\theta}(\epart{t + 1}{i}, Y_{t + 1})$;
        \EndFor
        \State \Return $\{(\epart{t + 1}{i}, \wgt{t + 1}{i})\}_{i = 1}^{\N}$
    \end{algorithmic}
\end{algorithm}

In the previous scheme, $\probdist(\{\wgt{t}{\ell}\}_{\ell = 1}^{\N})$ refers to the discrete probability distribution induced by the weights $\{ \wgt{t}{\ell} \}_{\ell = 1}^{\N}$. As a by-product, the historical trajectories of the particle filter provide jointly an estimate of the joint-smoothing distribution. These trajectories are constructed by linking up the particles with respect to ancestors. However, this method suffers from a well-known degeneracy phenomenon in the sense that the repeated resampling operations collapse the particle lineages as time increases. Consequently, the weighted empirical measures associated with the paths degenerate in the long run; see~\cite{olsson:cappe:douc:moulines:2006} for some discussion. 

A way to combat the degeneration is to use instead the backward decomposition presented above. Using the output of the bootstrap particle filter we obtain the particle approximation 
$$
    \bk{\post[part]{t; \theta}; \theta}(x_{t + 1}, \epart{t}{i}) = \frac{ \wgt{t}{i} \hd{\theta}(\epart{t}{i}, x_{t + 1})}{\sum_{\ell = 1} \wgt{t}{\ell} \hd{\theta}(\epart{t}{\ell}, x_{t + 1})} 
$$
of the backward kernel. Plugging this into the backward decomposition we arrive at the \emph{forward-filtering backward-smoothing} (FFBSm) algorithm, where $\post{0:T|T;\theta}(\testf)$ is approximated by
\begin{multline*}
    \post[part]{0:T | T; \theta}(\testf) = \sum_{i_0 = 1}^{\N} \cdots \sum_{i_T = 1}^{\N} \prod_{\ell = 0}^{T - 1} \frac{\wgt{\ell}{i_\ell} \hd{\theta}(\epart{\ell}{i_\ell}, \epart{\ell + 1}{i_{\ell + 1}})}{\sum_{\ell = 1}^{\N} \wgt{\ell}{\ell} \hd{\theta}(\epart{\ell}{\ell}, \epart{\ell + 1}{i_{\ell + 1}})} \\
    \times \frac{\wgt{T}{i_T}}{\wgtsum{T}}\testf(\epart{0}{i_0}, \ldots, \epart{T}{i_T}) \eqsp.
\end{multline*}
For a general objective function $\testf$, this occupation measure is impractical since the cardinality of its support grows geometrically fast with $T$. In the case where $\testf$ is of additive form the computational complexity is quadratic since the computation of the normalizing constants is required for each particle and each time step. Consequently, FFBSm is a computationally intensive approach.

In the case where the objective function is of additive form we can use the forward decomposition presented earlier to obtain an online algorithm; see~\cite{delmoral:doucet:singh:2009}. We denote the auxiliary functions $\{\tstattil{t}{i}\}_{i = 1}^{\N}$ initialized by setting $\tstattil{0}{i} = 0$ for all $i= \{1, \ldots, \N\}$. When a new observation is available, an update of the particle sample is followed by a recursive update of the auxiliary functions $\{ \tstattil{t}{i}\}_{i=1}^{\N}$ according to
\begin{equation} \label{eqn:bk:exp}
    \tstattil{t + 1}{i} = \sum_{j = 1}^{\N} \frac{\wgt{t}{j} \hd{\theta}(\epart{t}{j}, \epart{t+1}{i})}{\sum_{\ell = 1}^{\N} \wgt{t}{\ell} \hd{\theta}(\epart{t}{\ell}, \epart{t + 1}{i})} \{ \tstattil{t}{j} + \addf{t}(\epart{t}{j},\epart{t+1}{i}) \} \eqsp. 
\end{equation}
After this, the FFBSm estimate is formed as
$$
    \post[part]{0:t \mid t; \theta}(\af{t}) = \sum_{i = 1}^{\N} \frac{\wgt{t}{i}}{\wgtsum{t}}\tstattil{t}{i} \eqsp.
$$
Appealingly, this approach allows for online processing of the data and requires only the current particles and auxiliary statistics to be stored. Still, the computational complexity of the algorithm grows quadratically with the number of particles, since a sum of $N$ terms needs to be computed for each particle at each time step.

To speed up the algorithm, \cite{olsson:westerborn:2014:2} propose, in the \emph{particle-based rapid incremental smoother} (PaRIS) algorithm,  \eqref{eqn:bk:exp} to be replaced by a Monte Carlo estimate. Given $\{\tstat[i]{t - 1}\}_{i = 1}^{\N}$ we update the auxiliary statistic by drawing indices $\{\bi{t}{i}{\k}\}_{\k = 1}^{\K}$ according to the backward dynamics governed by the particle filter, i.e., drawing 
$$
 \bi{t}{i}{\k} \sim \probdist \left( \{ \wgt{t - 1}{\ell} \hd{\theta}(\epart{t - 1}{\ell},\epart{t}{i}) \}_{\ell = 1}^N \right) \eqsp.
$$
After this, each auxiliary statistic is updated through the Monte Carlo estimate
\begin{equation} \label{eqn:paris:upd}
\tstat[i]{t} = \K^{-1} \sum_{\k = 1}^{\K} \left( \tstat[\bi{t}{i}{\k}]{t - 1} + \addf{t}(\epart{t - 1}{\bi{t}{i}{\k}},\epart{t}{i}) \right), 
\end{equation}
and the estimate of $\post{0:t|t ;\theta}(\af{t})$ is obtained as $\sum_{i = 1}^{\N} \wgt{t}{i} \tstat[i]{t} / \wgtsum{t}$. Again, the procedure is initialized by setting $\tstat[i]{0} = 0$ for $i = \{1,\ldots, \N\}$. In this naive form the computational complexity is still quadratic; however, this approach can be furnished with an accept-reject trick found by \cite{douc:garivier:moulines:olsson:2010}, which reduces drastically the computational work. The accept-reject procedure can be applied when the transition density for the hidden chain is bounded, i.e., there exists some finite constant $\hd{\theta}^+$ such that $\hd{\theta}(x, x') \leq \hd{\theta}^+$ for all $(x, x') \in \set{X}^2$. This is a very weak assumption which is generally satisfied. In the scheme, an index proposal $J^*$ drawn from $\probdist(\{ \wgt{t - 1}{i} \}_{i = 1}^{\N})$ is accepted with probability $\hd{\theta}(\epart{t - 1}{J^*}, \epart{t}{i}) /\hd{\theta}^+$, and the procedure is repeated until acceptance. Under the \emph{strong mixing assumption} (see below for details) it can be shown that the expected number of proposals is \emph{bounded}; see~\cite{douc:garivier:moulines:olsson:2010,olsson:westerborn:2014:2}. Consequently, the overall computational complexity of PaRIS is \emph{linear}. In PaRIS $\K$ is a design parameter that should be at least 2 to achieve a stable algorithm; see \citet{olsson:westerborn:2014:2} for details. 

\citet{olsson:westerborn:2014:2} present a thorough theoretical study of the PaRIS algorithm. Under the weak assumption that the emission density is bounded and positive, it is established that the following central limit theorem holds true for all fixed $t$ and $\K$: as $\N \to \infty$,
$$
    \sqrt{\N} \sum_{i=1}^{\N}\frac{\wgt{t}{i}}{\wgtsum{t}}\left( \tstat[i]{t} - \post{0:t\mid t; \theta} (\af{t}) \right) \overset{\mathcal{D}}{\longrightarrow} \sigma_t(\af{t})Z,
$$
where $Z$ has standard Gaussian distribution and
$$
    \sigma_t^2(\af{t}) = \tilde \sigma_t^2(\af{t}) + \sum_{s = 0}^{t-1} \sum_{\ell = 0}^{s} \K^{\ell - (s + 1)} \delta_{\ell,s,t}^2,
$$
with $\tilde \sigma_t^2$ being the asymptotic variance of the FFBSm algorithm; see~\citet[Corollary 5]{olsson:westerborn:2014:2} for details. In the same work, the numerical stability of PaRIS is studied under the strong mixing assumptions
\begin{itemize}
    \item[(i)] $0 < q_{\theta}^{-} \leq \hd{\theta}(x,\tilde x) \leq q_{\theta}^+$ for all $(x, \tilde x) \in \set{X}^2$,
    \item[(ii)] $\|\md{\theta}(\cdot, Y_t) \|_{\infty} \leq \mdup_{\theta}$ and $0 < \mdlow_{\theta} \leq \int \md{\theta}(\tilde x, Y_t) \hd{\theta}(x, \tilde x) \rmd \tilde x$ for all $x, t$,
\end{itemize}
which are standard in the literature and point to applications where the state space $\set{X}$ is compact. Moreover, it is assumed that there exists a constant $| \addf{} |_{\infty}$ such that for all $t$, $\oscn(\addf{t}) \leq | \addf{} |_{\infty} < \infty$. Under these assumptions it is shown that there exist constants $c$ and $\tilde{c}$ such that, for $\K \geq 2$,
$$
    \limsup_{t \to \infty} \frac{1}{t} \sigma_t^2(\af{t}) \leq | \addf{} |_{\infty}^2 \left( c + \frac{\tilde c}{\K - 1} \right).
$$
This bound implies that the Monte Carlo error added by PaRIS at each time step is uniformly bounded in time, which shows that the algorithm stays numerically stable in the long run. Moreover, as the previous bound is inversely proportional to $\K$, we may draw the conclusion that $\K$ should be kept at a moderate value, say, less than 10. Consequently, the asymptotic variance is of order $\mathcal{O}(t)$, which is the best possible for a Monte Carlo approximation on the path space; see \citet[Theorem~8]{olsson:westerborn:2014:2} for details. See also \citet{delmoral:guionnet:2001} and \citet{douc:garivier:moulines:olsson:2010} for similar stability results for the standard particle filter and the FFBSm algorithm, respectively.



Now, we may cast the PaRIS algorithm into the framework of the online EM algorithm of \cite{cappe:2009} by simply replacing \eqref{eqn:paris:upd} by the the updating formula 
$$
    \tstat[i]{t} = \K^{-1} \sum_{\k = 1}^{\K} \left( (1 - \gamma_t) \tstat[\bi{t}{i}{\k}]{t - 1} + \gamma_t \addf{t - 1}(\epart{t - 1}{\bi{t}{i}{\k}},\epart{t}{i}) \right) \eqsp,
$$
where again the sequence $\{\gamma_t\}_{t \in \nset}$ should satisfy the usual stochastic approximation requirements. A standard choice is to set $\gamma_t = t^{-\alpha}$ for $.5 < \alpha \leq 1$. The algorithm is summarized in Algorithm~\ref{alg:PaRIS:OEM}. 

\begin{algorithm}[htb]
    \caption{PaRIS-based online EM}
    \label{alg:PaRIS:OEM}
    \begin{algorithmic}[1]
        \State set $\tstat[i]{0} = 0$ for $i =  \{1, \ldots, \N\}$;
        \State under some $\theta_0$, generate $\{(\epart{0}{i}, \wgt{0}{i})\}_{i = 1}^{\N}$ targeting $\post{0; \theta_0}$;
        \For{$t = 1 \to T$}
        \State run $\{ (\epart{t}{i}, \wgt{t}{i}) \}_{i = 1}^\N \gets \PF( \{ (\epart{t-1}{i}, \wgt{t-1}{i}) \}_{i = 1}^\N; \theta_{t-1} )$;
        \For{$i = 1 \to \N$}
            \For{$\k = 1 \to \K$}
                \State draw $\bi{t}{i}{j} \sim \probdist( \{ \wgt{t-1}{\ell} \hd{\theta_{t-1}}(\epart{t-1}{\ell}, \epart{t}{i} ) \}_{\ell = 1}^\N )$;
            \EndFor
            \State set 
            $$\tstat[i]{t} \gets \K^{-1} \sum_{\k = 1}^{\K} \left( (1 - \gamma_t) \tstat[\bi{t}{i}{\k}]{t-1} + \gamma_t \term{t-1}(\epart{t-1}{\bi{t}{i}{\k}}, \epart{t}{i}) \right);
            $$
            \State set $\displaystyle \theta_t \gets \Lambda \left( \sum_{i=1}^{\N} \frac{\wgt{t}{i}}{\wgtsum{t}} \tstat[i]{t} \right)$;
        \EndFor
        \EndFor
        \State \Return $\theta_T$.
    \end{algorithmic}
\end{algorithm}

As mentioned, the standard batch EM algorithm update, at iteration $i + 1$, the parameters using 
$$
    t^{-1} \post{0:t|t;\theta_{i}}\af{t} = t^{-1}\E_{\theta_i} \left[ \sum_{\ell = 0}^{t - 1} \addf{\ell}(X_\ell, X_{\ell + 1}) \mid Y_{0:t} \right] \eqsp,
$$
and it can be established, under additional assumptions, that every fixed point of EM is indeed a stationary point of the likelihood. The online EM algorithm updates instead, at iteration (i.e., time step) $t$, the parameters on the basis of
\begin{multline*}
    \E_{\theta_{0:t - 1}} \left[ \gamma_t \addf{t - 1}(X_{t - 1}, X_t) + (1 - \gamma_t) \sum_{\ell = 1}^{t - 1} \right. \\
    \left. \times \left(\prod_{i = \ell + 1}^{t - 1} (1 - \gamma_i) \right) \gamma_\ell \addf{\ell - 1}(X_{\ell - 1}, X_\ell) \mid Y_{0:t} \right] \eqsp,
\end{multline*}
where the subscript $\theta_{0:t-1}$ indicates that the expectation is taken under the model dynamics governed by $\theta_i$ at time $i + 1$. Thus, if the sequence $\{Ê\theta_t \}_{t \in \nset}$ converges, we may expect, since the factor $\prod_{i = \ell + 1}^{t - 1} (1 - \gamma_i)$ reduces the influence of early parameter estimates, that a fixed point of the online EM procedure coincides with a stationary point of the asymptotic contrast function, i.e., by identifiability, the \emph{true} parameter value (see \cite{douc:olsson:moulines:vanhandel:2011}).  At present, a convergence result for the online EM algorithm for SSMs is lacking \citep[a theoretical discussion is however given by][]{cappe:2009}, but for independent observations (e.g., mixture models) convergence is shown in~\cite{cappe:moulines:2009}. 

Another algorithm that is worth mentioning here is the \emph{block online EM algorithm} \citep{lecorff:fort:2013}, where the parameter is only updated at fixed and increasingly separated time points. This algorithm can be shown to converge; however, simulations indicate that this block processing approach is less advantageous than updating the parameter at every time step. An overview of parameter estimation methods is given by~\cite{kantas:doucet:singh:maciejowski:chopin:2014}.

\section{Simulations}

We test the algorithm on two different models, the linear Gaussian model in Example~\ref{ex:lg} and a \emph{stochastic volatility model}. With these simulations we wish to show that the PaRIS-based algorithm is preferred to the FFBSm-based version. In the implementations we start updating the parameter first after a few observations have been processed in order to make sure that the filter estimates are stable.

\subsection{Linear Gaussian model}

For the linear Gaussian model in Example~\ref{ex:lg} we compare the parameter estimates produced by the PaRIS-based algorithm with those produced by the FFBSm-based algorithm of~\cite{delmoral:doucet:singh:2009}. The observed data are generated by simulation under the parameters $\theta = (.8, .4^2, .9^2)$. We tune the number of particles of both algorithms and the number of backward draws $\K$ in the PaRIS-based algorithm such that the computational time of both algorithms are similar. This implies $250$ particles for the FFBSm-based algorithm and 1250 particles and $\K = 5$ for the PaRIS-based algorithm. We also restrict ourselves to estimation of the parameters $a$ and $\sigma_V$.

\begin{figure}
    \centering
    \begin{subfigure}[t]{0.45\textwidth}
        \centering
        \includegraphics[width=\textwidth]{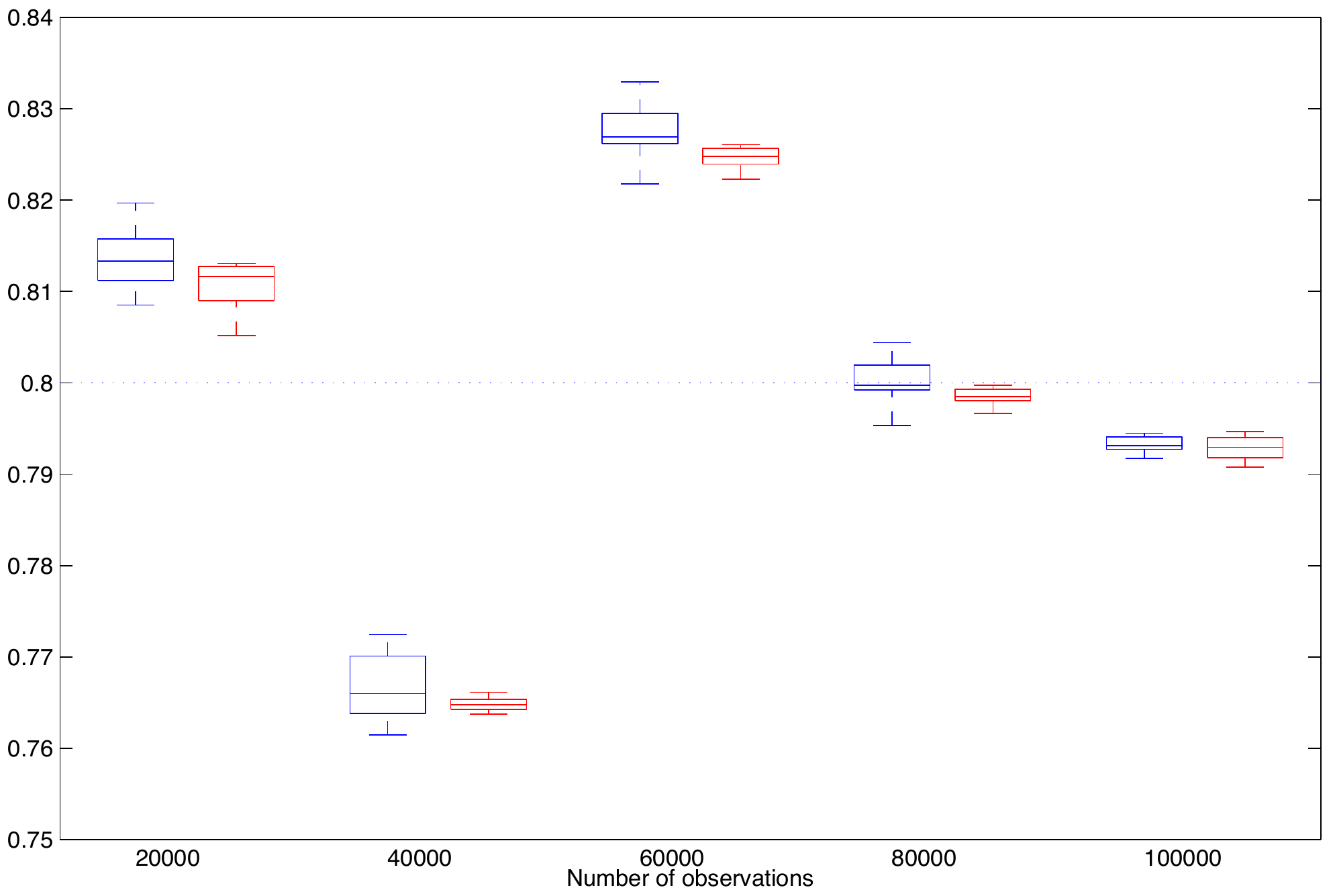}
        \caption{Estimation of $a$}
    \end{subfigure}
    \begin{subfigure}[t]{0.45\textwidth}
        \centering
        \includegraphics[width=\textwidth]{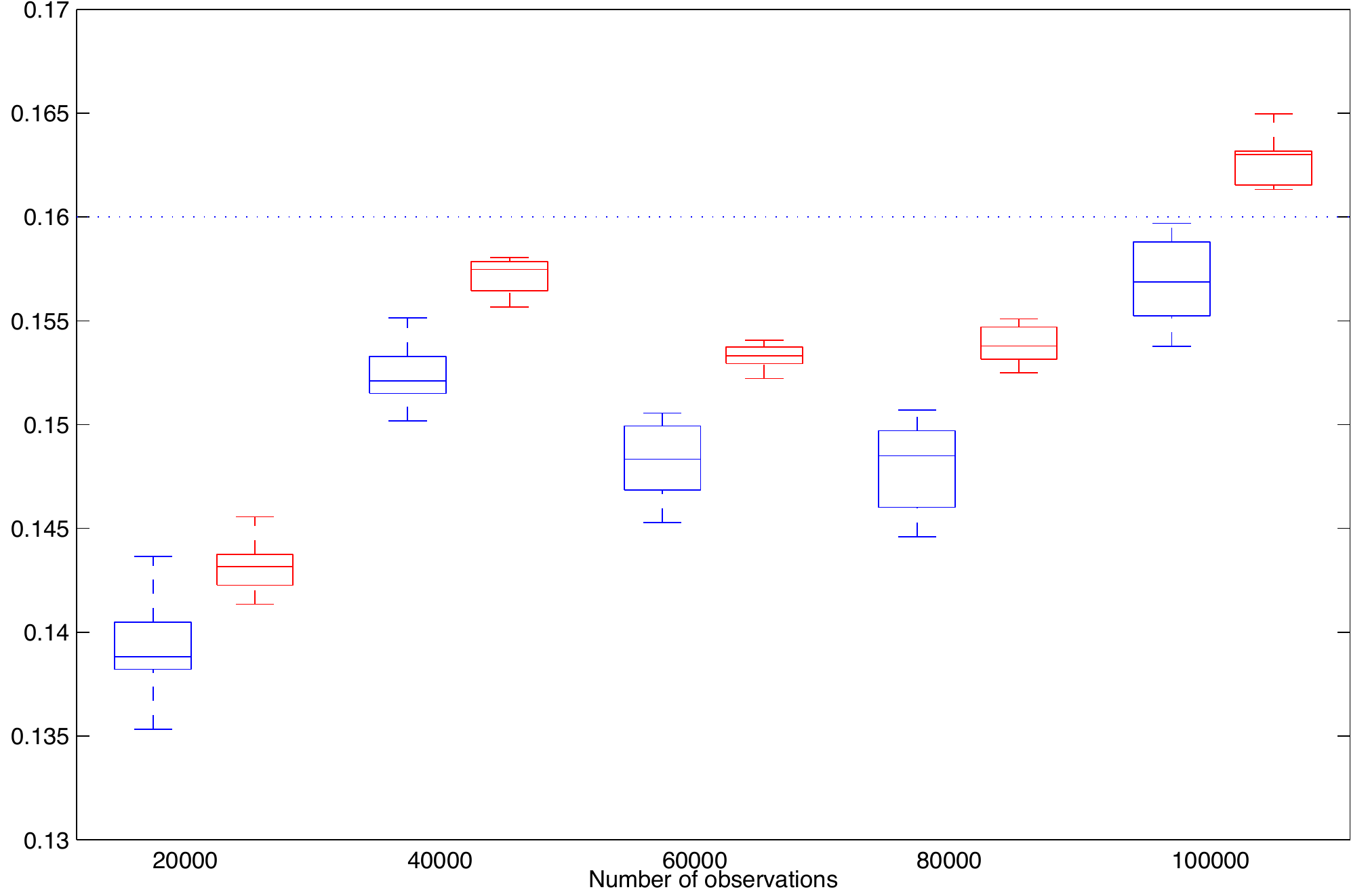}
        \caption{Estimation of $\sigma_V^2$}
    \end{subfigure}
    \caption{FFBSm-based (left boxes) and PaRIS-based estimates (right boxes) of $a$ (panel a) and $\sigma_V^2$ (panel b) in the linear Gaussian model. Dashed horizontal lines indicate true parameter values.}
    \label{fig:lg}
\end{figure}

In Figure~\ref{fig:lg} we present output of the algorithms based on 10 independent runs on the same simulated data, where $\theta_0 = (.1, 2^2, .9^2)$ and where we update only the $a$ and $\sigma_V^2$ parameters. We set $\gamma_t = t^{-0.6}$ and start updating the parameters after 60 steps. As clear form the plot, both algorithms tend towards the true parameters. In addition, the PaRIS-based algorithm exhibits, as a consequence of the larger particle sample size, a lower variance. 

\subsection{Stochastic volatility model}
The stochastic volatility model is given by 
\[
    \begin{split}
        X_{t + 1} &= \phi X_t + \sigma V_t \eqsp, \\
        Y_t &= \sigma \exp(X_t / 2) U_t \eqsp,
    \end{split}
    \quad t \in \nset \eqsp,
\]
where again $\{ÊV_t \}_{t \in \nset}$ and $\{ÊU_t \}_{t \in \nset}$ are independent sequences of mutually independent standard Gaussian noise variables. The parameters of the model are $\theta = (\phi, \sigma^2, \beta^2)$, the sufficient statistics are given by
\begin{align*}
    \sadd{t}{1}(x_{t:t+1}) &= x_t^2, & \sadd{t}{2}(x_{t:t+1}) &= x_t x_{t+1}, \\ 
    \sadd{t}{3}(x_{t:t+1}) &= x_{t+1}^2, & \sadd{t}{4}(x_{t:t+1}) &= y_{t+1}^2 \exp(-x_{t+1}),
\end{align*}
and parameter updates are given by 
\begin{align*}
    \Lambda(z_1, z_2, z_3, z_4) = \left( \frac{z_2}{z_3}, z_3 - \frac{z_2^2}{z_1}, z_4  \right) \eqsp.
\end{align*}
We generate the data through simulation using the parameters $\theta = (.975, .16^2, .63^2)$. Again, we set the parameters of the algorithm in such a way that the computational times of the two algorithms are similar. This implies $110$ and $500$ particles for the FFBSm-based and PaRIS-based algorithms, respectively. In addition, the latter used $\K = 4$ backward draws. 

\begin{figure}
    \centering
    \begin{subfigure}[t]{0.45\textwidth}
        \centering
        \includegraphics[width=\textwidth]{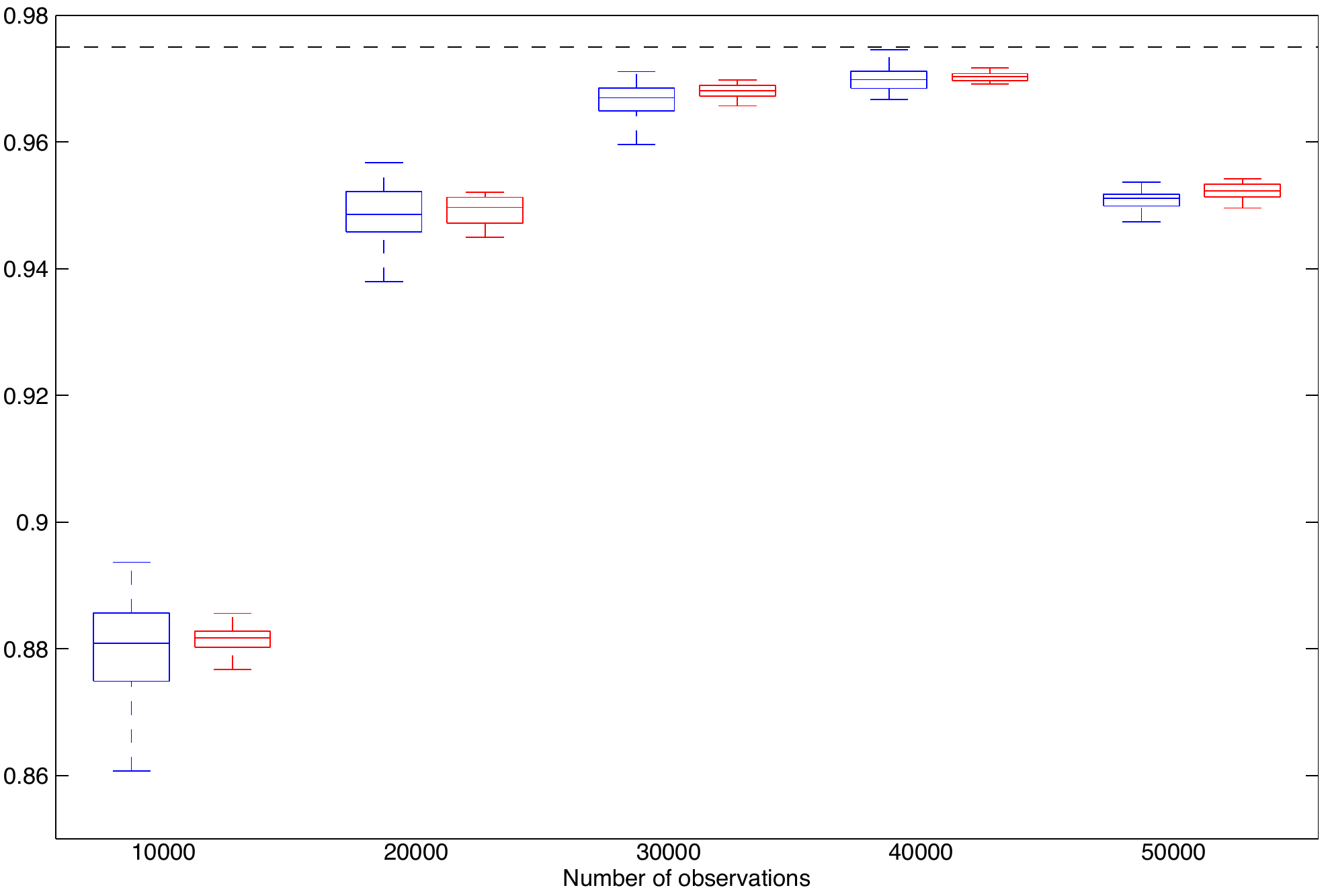}
        \caption{Estimation of $\phi$}
    \end{subfigure}
    \begin{subfigure}[t]{0.45\textwidth}
        \centering
        \includegraphics[width=\textwidth]{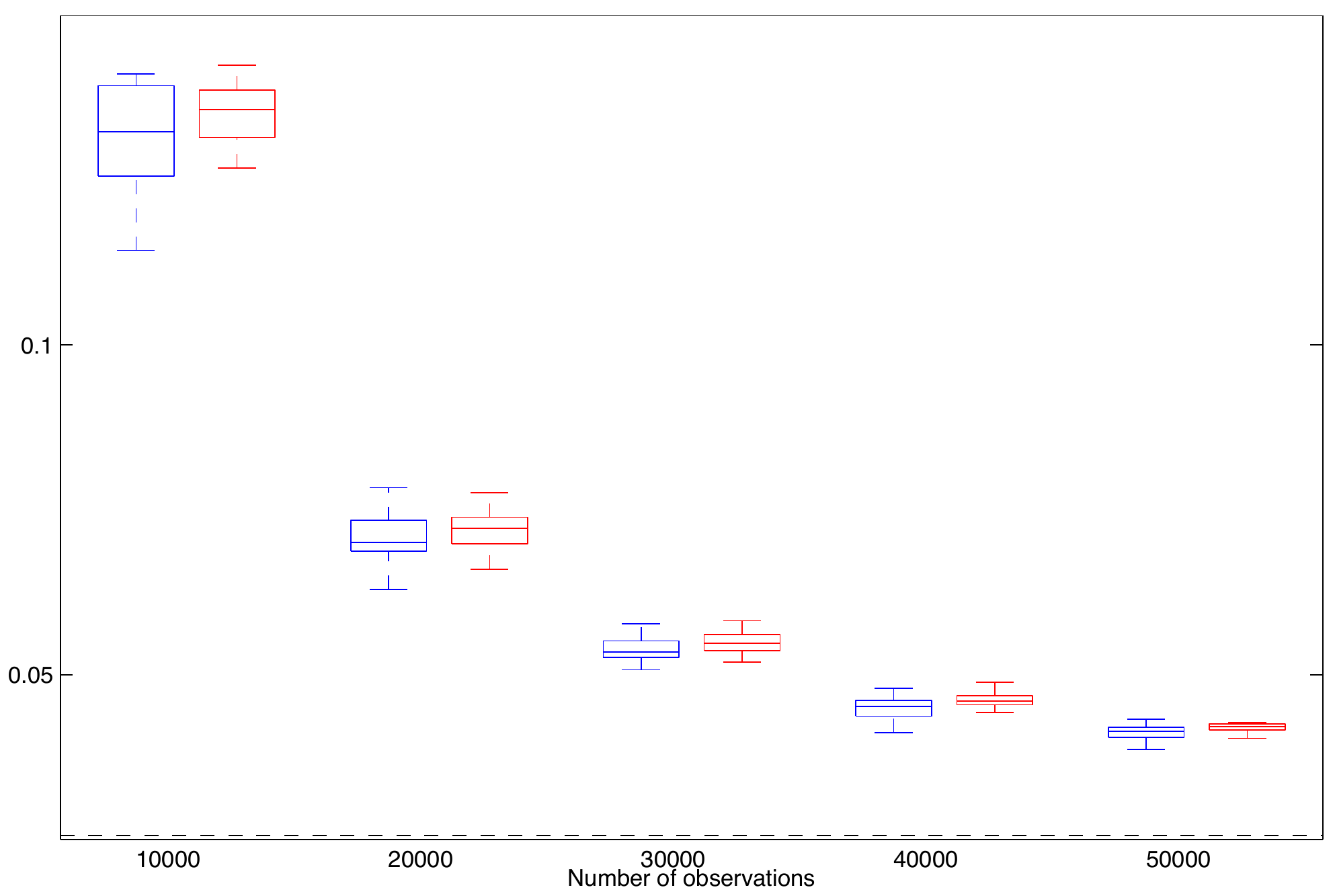}
        \caption{Estimation of $\sigma^2$}
    \end{subfigure}
    \begin{subfigure}[t]{0.45\textwidth}
        \centering
        \includegraphics[width=\textwidth]{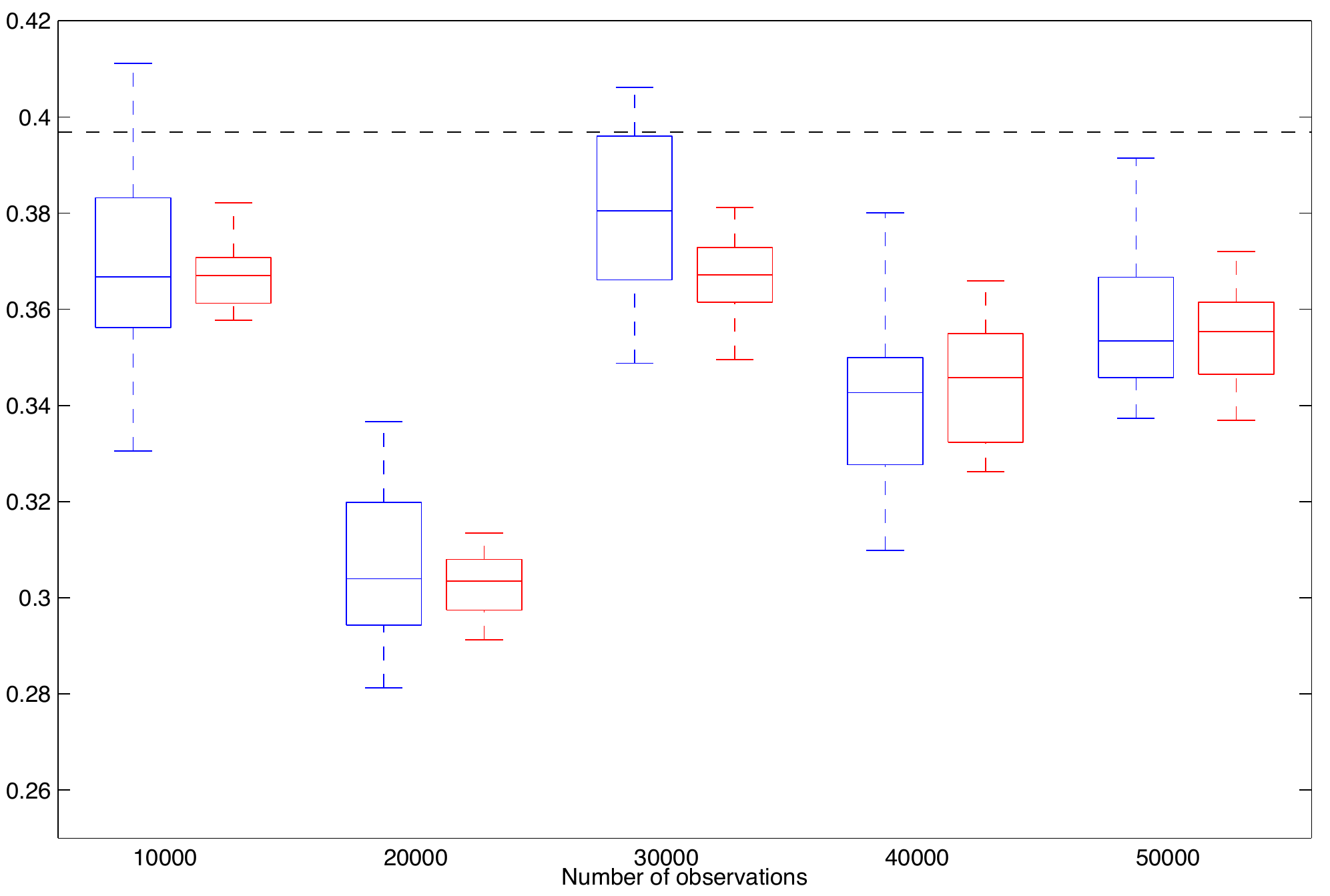}
        \caption{Estimation of $\beta^2$}
    \end{subfigure}
    \caption{FFBSm-based (left boxes) and PaRIS-based estimates (right boxes) of $\phi$ (panel a), $\sigma^2$ (panel b), and $\beta^2$ (panel c) in the stochastic volatility model. Dashed horizontal lines indicate true parameter values.}
    \label{fig:sv}
\end{figure}

In Figure~\ref{fig:sv} we present the output of both algorithms from 20 independent runs using the same data input for each run. We initialize the algorithms with $\theta_0 = (.5, .8^2, 1^2)$, use $\gamma_t = t^{-.6}$, and start the parameter updating after 60 observations. We notice, as for the previous model, that both algorithms seem to converge towards the correct parameters and that the FFBSm-based algorithm exhibits the higher variance. 

\begin{figure}
    \centering
    \includegraphics[width=0.45\textwidth]{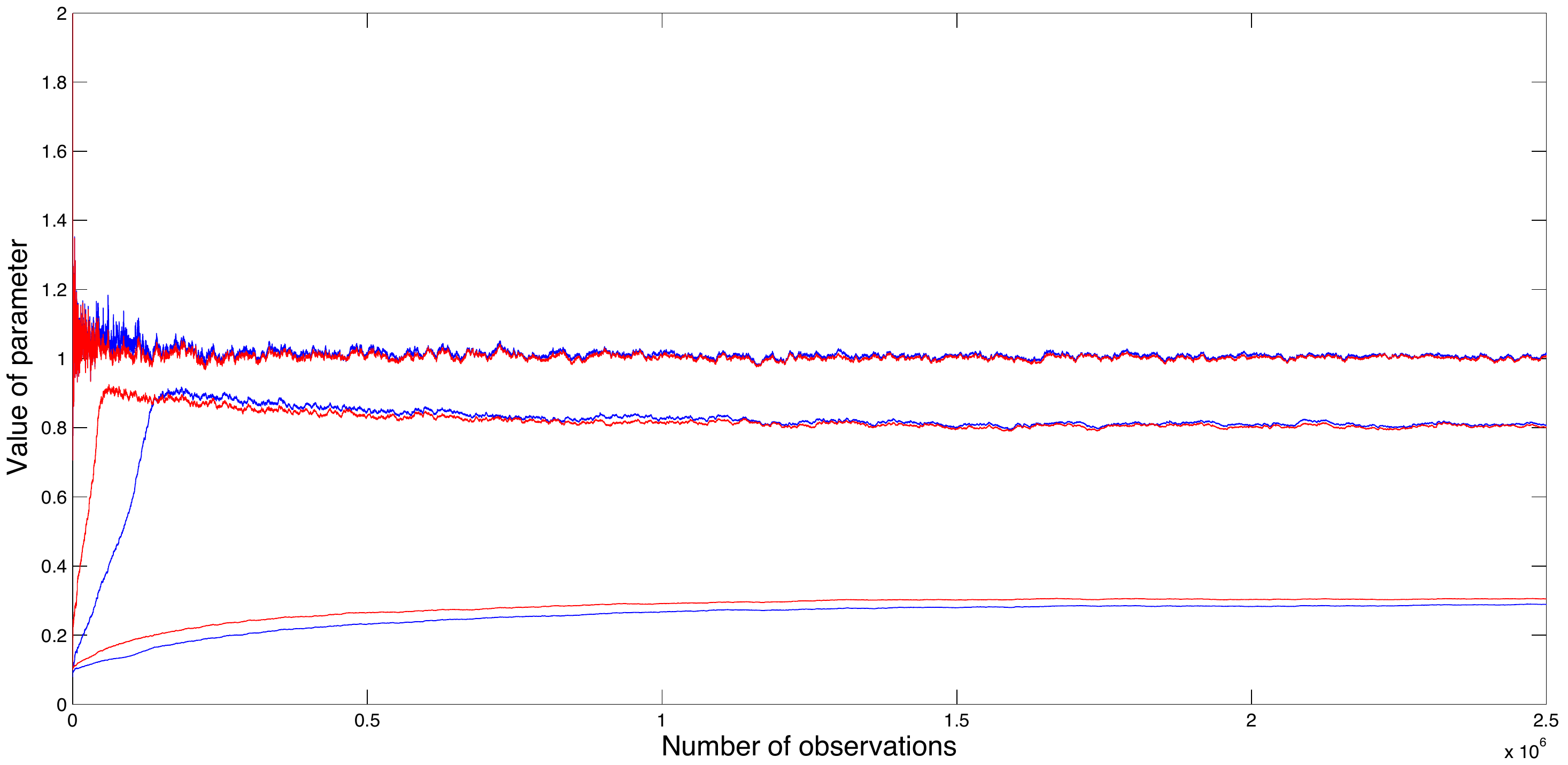}
    \caption{Estimation of the parameters in the stochastic volatility model using $2,\!500\!000$ observations. The red lines correspond to the PaRIS-based estimates and the blue lines correspond to the FFBSm-based estimates. For bottom to top the parameters are $\sigma$, $\phi$, and $\beta$, respectively.}
    \label{fig:sv:long}
\end{figure}

Finally, to show that the algorithm indeed converges we perform one run of the algorithms on the stochastic volatility model parameterized by $\theta = (.8, .1, 1)$ using $\theta_0 = (.1, .1^2, 2^2)$ and as many as $T = 2,\!500,\!000$ observations. For the FFBSm-based algorithm we used $\N = 125$ particles and for the PaRIS-based algorithm we used $\N = 500$ and $\K = 2$. Both algorithms use $\gamma_t = t^{- .6}$ and do not update the parameters for the first 60 observations. The results are reported in Figure~\ref{fig:sv:long}, which indicates convergence for both algorithms. Taking the mean of the last 1000 parameter estimates yields the estimates $(.802, .093, 1.01)$Êand $(.807, .084, 1.03)$ for the PaRIS-based and FFBSm-based algorithms, respectively. 

\section{Conclusions}

We have presented a new particle-based version of the online EM algorithm for parameter estimation in general SSMs. This new algorithm, which can be viewed as a hybrid of the PaRIS smoother proposed by \cite{olsson:westerborn:2014:2} and the online EM algorithm of \cite{cappe:2009}, has a computational complexity that grows only linearly with the number of particles, which results in a fast algorithm. Compared to existing algorithms, this allows us to use considerably more particles and, consequently, produce considerably more accurate estimates for same amount of computational work.

\bibliography{biblio}

\end{document}

%% file: ow_sysid2015_def.tex
\newcommand{\af}[1]{s_{#1}} 

\newcommand{\addf}[1]{\termletter_{#1}} 

\newcommand{\bi}[3]{J_{#1}^{(#2, #3)}}
\newcommandx{\bk}[2][1=]{ 
\ifthenelse{\equal{#1}{}}
{\overleftarrow{q}_{#2}}
{\overleftarrow{\kernel{Q}}_{#2}^{#1}}
}

\newcommand{\E}{\mathbb{E}}
\newcommand{\epart}[2]{\xi_{#1}^{#2}}
\newcommand{\eqsp}{}

\newcommand{\hd}[1]{q_{#1}} 

\newcommand{\kernel}[1]{\mathbf{#1}}
\newcommand{\kletter}{\tilde{\N}}
\renewcommand{\k}{j}
\newcommandx{\K}[1][1=]{
\ifthenelse{\equal{#1}{}}{{\kletter}}{{\tilde{\N}^{#1}}}
}

\newcommand{\md}[1]{g_{#1}} 
\newcommand{\mdlow}{\ushort{\delta}}
\newcommand{\mdup}{\bar{\delta}}

\newcommand{\N}{N}
\newcommand{\nset}{\mathbb{N}}

\newcommand{\PF}{\mathsf{PF}}
\newcommandx\post[2][1=]{
\ifthenelse{\equal{#1}{}}
	{\phi_{#2}}
	{\phi_{#2}^\N}
}
\renewcommand{\prob}{\mathbb{P}} 
\newcommand{\probdist}{\mathsf{Pr}}

\newcommand{\oscn}{\operatorname{osc}}
\newcommand{\rmd}{\mathrm{d}}

\newcommand{\set}[1]{\mathsf{#1}} 

\newcommand{\sadd}[2]{\tilde{s}_{#1}^{#2}}
\newcommand{\statistic}[1]{\tau_{#1}}
\newcommand{\term}[1]{\termletter_{#1}}
\newcommand{\termletter}{\tilde{s}}
\newcommand{\testf}{f}
\newcommandx\tstat[2][1=]{
\ifthenelse{\equal{#1}{}}
	{\tstatletter_{#2}}
	{\tau_{#2}^{#1}}
}
\newcommand{\tstattil}[2]{\tilde{\tau}_{#1}^{#2}}

\newcommand{\wgt}[2]{\omega_{#1}^{#2}}
\newcommand{\wgtsum}[1]{\Omega_{#1}}

\newcommand{\Xinit}{\chi}
